\documentclass[twocolumn,10pt]{IEEEtran}
\usepackage{algorithm,algorithmic,amsbsy,amsmath,amssymb,epsfig,subfig,bbm,mathrsfs,mathdots,arydshln,arydshln,multirow,verbatim,booktabs,setspace,url,stfloats,fixltx2e,cite,nomencl,epstopdf,caption,xcolor}

\newcommand{\beq}{\begin{equation}}
\newcommand{\eeq}{\end{equation}}
\newcommand{\bitm}{\begin{itemize}}
\newcommand{\ba}{\begin{array}}
\newcommand{\ea}{\end{array}}
\newcommand{\eitm}{\end{itemize}}
\newcommand{\beqn}{\begin{eqnarray}}
\newcommand{\eeqn}{\end{eqnarray}}
\newcommand{\beqno}{\begin{eqnarray*}}
\newcommand{\bma}{\begin{displaymath}}
\newcommand{\ema}{\end{displaymath}}
\newcommand{\bnu}{\begin{enumerate}}
\newcommand{\enu}{\end{enumerate}}
\newcommand{\bce}{\begin{center}}
\newcommand{\ece}{\end{center}}
\newcommand{\btb}{\begin{tabular}}
\newcommand{\etb}{\end{tabular}}

\begin{document}

\title{Evolutionary Carrier Selection for Shared Truck Delivery Services}
\author{\IEEEauthorblockN{Rakpong Kaewpuang$^{\mathrm{1}}$, Suttinee Sawadsitang$^{\mathrm{2}\ast}$, Dusit Niyato$^{\mathrm{1}}$, and Han Yu$^{\mathrm{1}}$ \\
\IEEEauthorblockA{$^{\mathrm{1}}$School of Computer Science and Engineering, Nanyang Technological University \\
                  $^{\mathrm{2}}$College of Art, Media and Technology, Chiang Mai University                 
} } 
\thanks{ $^{\ast}$Corresponding author (e-mail: suttinee.s@cmu.ac.th).}
}
\maketitle

\begin{abstract}
With multiple carriers in a logistics market, customers can choose the best carrier to deliver their products and packages. In this paper, we present a novel approach of using the stochastic evolutionary game to analyze the decision-making of the customers using the less-than-truckload (LTL) delivery service. We propose inter-related optimization and game models that allow us to analyze the vehicle routing optimization for the LTL carriers and carrier selection for the customers, respectively. The stochastic evolutionary game model incorporates a small perturbation of customers' decision-making which exists due to irrationality. The solution of the stochastic evolutionary game in terms of stochastically stable states is characterized by using the Markov chain model. The numerical results show the impact of carriers' and customers' parameters on the stable states.
\end{abstract}


\section{Introduction}

Product and package delivery management is a classical but important problem in the logistics industry. Especially, with the tremendous popularity of online shopping, package delivery demand increases exponentially. Together with fierce competition among logistics businesses to attract customers, various challenging issues emerge from a classical vehicle routing problem (VRP). Specifically, instead of just focusing on finding optimal vehicle routes, e.g., minimizing traveling costs, competition and cooperation must be considered when considering logistics resource/vehicle management. Game theory becomes a suitable mathematical tool to analyze various logistics scenarios in which each logistics company can adapt its strategy to reach the best payoff, for example, defined in terms of utility or profit, given the strategies of other companies. Some applications of game theory in logistics exist, e.g.,~\cite{farokhi2013}~\cite{aswani2011}. In the logistics industry, multiple carriers own vehicles/trucks for package delivery. Customers can freely choose the carrier to meet their demands and requirements. While the major factor in deciding on the carrier to use is a price or service fee, recently performance and reliability of the delivery service become increasingly concerning. For example, customers may not always choose the cheapest carrier if their package delivery is time-sensitive. Therefore, customers can rationally select the carrier based on their preference, defined as a utility. The aforementioned scenario can be generally called the carrier selection problem. Again, the game theory appears to be a promising approach for analyzing and obtaining the solution of this problem. However, due to a large number of customers, the standard noncooperative game, which is tractable only for a few players, may not be suitable, and we have to resort to the other game model for solving the carrier selection problems of the customers.

In this paper, we focus on the carrier selection problem in the less-than-truckload (LTL) environment where a small freight or freight that does not require a full space of a vehicle is delivered~\cite{c-w-chu-ejor2005-heuristic}. The customers, i.e., players, can adaptively select LTL carriers\footnote{In the rest of the paper, we use ``LTL carrier'' and ``carrier'' interchangeably.} to achieve the highest payoff given the decision of the other customers. The payoff depends on the delivery performance, i.e., delay, which is again related to the delivery route. Considering this fact, we introduce the stochastic evolutionary game to analyze the carrier selection problem. The evolutionary game is applied as it has been shown to be tractably applicable for a large number of players. 

Additionally, a small perturbation due to the irrational decision-making of the players can be also captured in the model. The game model can show the impact of the number of customers choosing a certain carrier on the delivery performance, i.e., average delay. Moreover, the impact of the physical parameter, i.e., the location of a carrier, on the stochastically stable state of the game is presented. 

The major contributions of this paper can be summarized as follows:  
\begin{itemize}
    \item We propose inter-related optimization and game models to analyze the vehicle routing optimization for the LTL carriers and carrier selection for the customers, respectively.
    \item In the proposed models, we formulate and solve the integer programming  (IP) model to obtain the minimum total traveling cost of carriers. We apply the stochastic evolutionary game model to achieve stable states that customers obtain the minimum delivery delay and service fee.      
   \item We show the superiority of the proposed models by evaluating the performance with a real trace data and comparing the solution of the proposed models with the baseline model.   
\end{itemize} 

\section{Related Work}
\label{sec:relatedwork}

Optimizing the performance of package delivery from a depot or warehouse to customers and vice versa, e.g., to minimize the cost and delay, is a classical problem known as a vehicle routing problem with private fleet and common carrier (VRPPC)~\cite{c-w-chu-ejor2005-heuristic}. For example, the author in~\cite{c-w-chu-ejor2005-heuristic} proposed a truckload and less-than-truckload heuristic algorithm to solve the VRPPC. The objective of the algorithm is to minimize the total transportation cost by selecting private and outsourcing trucks for delivery. To tackle the problems, various heuristic algorithms were proposed. However, the carriers have limited resources in terms of vehicles and drivers. Game theory has been applied to address various issues in logistics. For example, in~\cite{farokhi2013}, the authors presented the routing game in a heterogeneous environment, i.e., the players as drivers and vehicles. The players choose the number of flows to reach the Nash equilibrium. A similar flow game was analyzed in~\cite{aswani2011} in which the impact of GPS information available to the drivers is taken into account.

However, the carrier selection problem by customers to deliver their packages with a minimum delay has not been studied before. Especially, when the customers can adjust their decisions adaptively considering the decisions of other customers in the population. The evolutionary game is a promising tool for analyzing this problem, and thus it is the main focus of this paper.

\section{System Model and Assumptions}
\label{sec:systemmodel}

\begin{figure}[htb]
 \centering
 \captionsetup{justification=centering}
  \subfloat[LTL carrier selection.]{\label{fig:systemmodel}\includegraphics[width=0.24\textwidth]{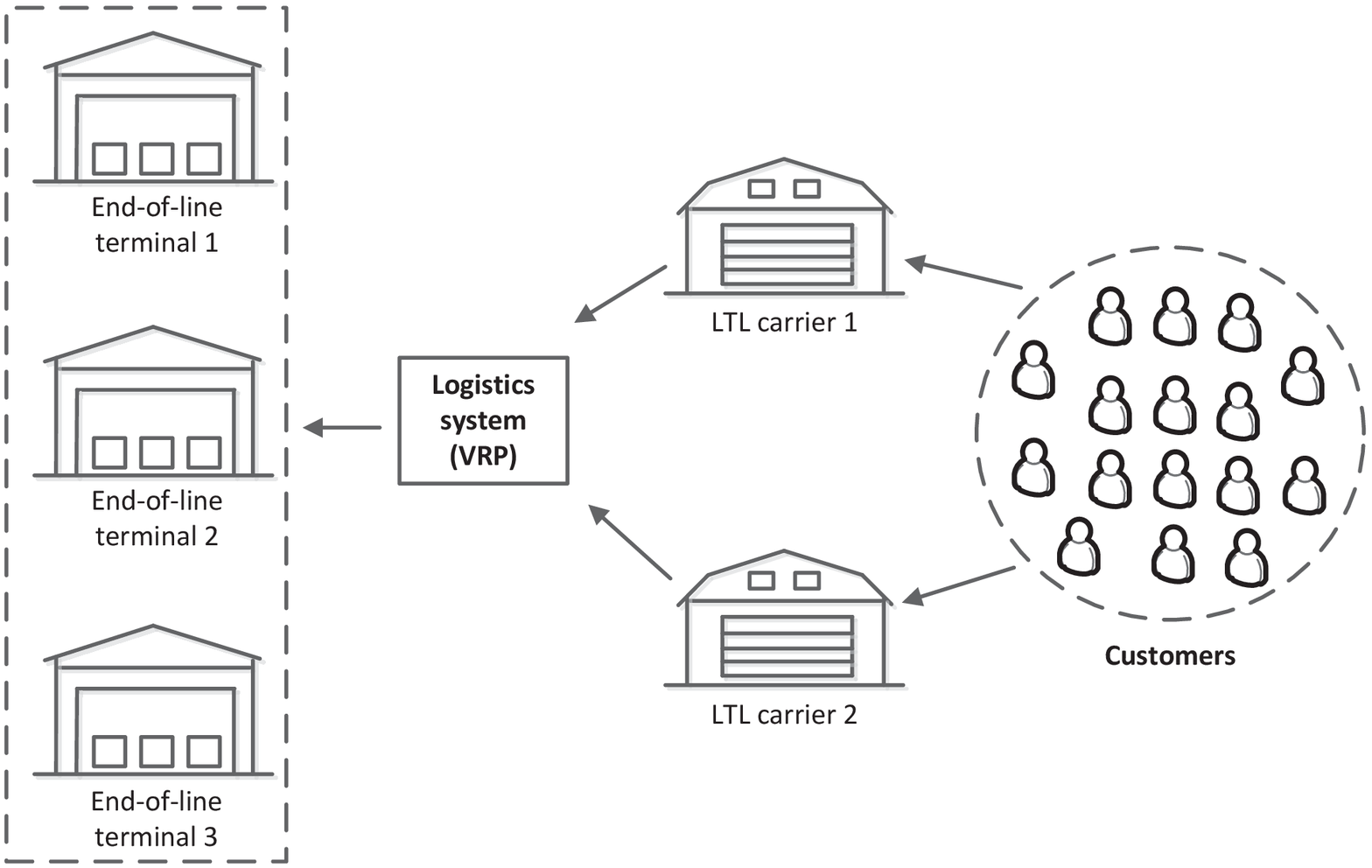}} 
  \subfloat[Reference locations.]{\label{fig:vp-c2-w-dist}\includegraphics[width=0.24\textwidth]{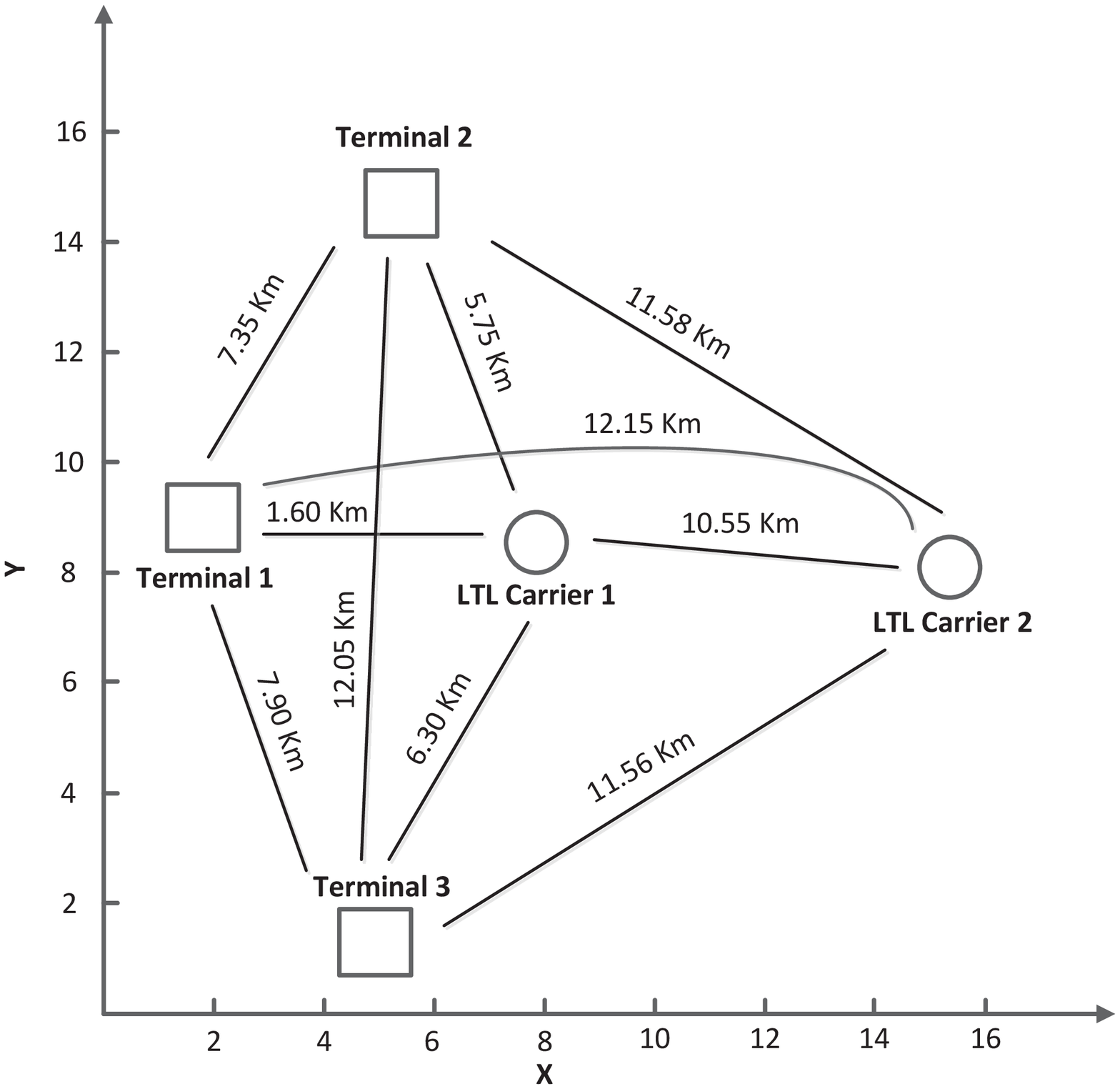}} 
 \caption{(a) LTL carrier selection and (b) Reference locations of two carriers and three terminals.}
 \label{fig:systemmodel-vp-c2-w-dist}
 \vspace{-0.2cm}
\end{figure} 

We consider the carrier selection scenario that the customers choose an LTL carrier to deliver packages to the terminals, which is illustrated in Fig. \ref{fig:systemmodel-vp-c2-w-dist}(a). In particular, the carrier collects packages from customers, e.g., from a collection point. Then the carrier loads the packages into an LTL truck and delivers the packages to the terminals. The carrier can optimize the delivery of the packages, which can be considered as the vehicle routing problem (VRP) with the fixed number of trucks. The objective is to minimize the delivery cost for the carrier. Alternatively, there are multiple carriers, and the customers select the carrier based on their (dis)utility which depends on service fee and delivery delay. Specifically, the customers prefer the carrier with the smallest fee and delay.

For the carrier selection, we define sets and notations as follows. Let $\mathcal{T}$ denote the set of terminals that LTL carriers deliver packages to, i.e., $\mathcal{T} = \{1,2,3,\dots,T\}$ where $T$ is the maximum number of terminals. Let $\mathcal{N}$ denote the set of LTL carriers, i.e., $\mathcal{N} = \{1,2,3,\dots,N \}$ where $N$ is the maximum number of LTL carriers. Let $\mathcal{S}$ denote the set of customers, i.e., $\mathcal{S} = \{1,2,3,\dots,S\}$ where $S$ is the maximum number of customers. In the carrier selection scenario, we are interested in analyzing how the customers choose the carrier. Therefore, in the following sections, we introduce the vehicle routing optimization model for the carriers and the stochastic evolutionary game model for the customers. Given the number of customers choosing a certain carrier, the vehicle routing optimization model is used to obtain the optimal route and the corresponding delivery delay of the carrier. Given the delivery delay, the stochastic evolutionary game model is applied to obtain the decision of customers selecting any carrier.


\section{Vehicle Routing Optimization} 
\label{sec:il-formulation}

We propose the IP model for vehicle routing optimization for a certain LTL carrier. The set of vehicles of the carrier is denoted by $\mathcal{V}$ and the set of terminals to be visited by the vehicles is denoted by $\mathcal{L}$. The IP model is presented in (\ref{ilf-obj})-(\ref{ilf-cons12}). The objective function in (\ref{ilf-obj}) is to minimize the total traveling cost of the carrier. 

\vspace{-0.2cm}
\begin{figure}[htb] 
\beqn
	&& \min_{x_{v},y_{i,j,v},z_{i,v}} \sum_{ v \in \mathcal{V} } x_{v} C^{\mathrm{inc}}_{v}  \label{ilf-obj} \\ \nonumber 
	&&	 + \sum_{v \in \mathcal{V}} \sum_{i \in \mathcal{L}} \sum_{j \in \mathcal{L}, i \neq j} y_{i,j,v} C^{\mathrm{tvc}}_{i,j,v} +  \sum_{v \in \mathcal{V}} \sum_{i \in \mathcal{U}} z_{i,v} C^{\mathrm{mcn}}	,	 \\ \nonumber
	&&	\mbox{subject to} \\  
	&& \sum_{j \in \mathcal{L}} y_{i,j,v} = z_{i,v}, \forall i \in \mathcal{U}, \forall v \in \mathcal{V}, \label{ilf-cons1} \\
	&& \sum_{j \in \mathcal{L}} y_{j,i,v} = z_{i,v}, \forall i \in \mathcal{U}, \forall v \in \mathcal{V}, \label{ilf-cons2} \\
	&& \sum_{ k \in \mathcal{V}} z_{0,v} \leq |\mathcal{V}|, \label{ilf-cons3} \\
	&&  \sum_{v \in \mathcal{V}} \sum_{i \in \mathcal{L}, i \neq j} y_{i,j,v}  = 1 , \forall j \in \mathcal{U}, \label{ilf-cons4} \\
	&&  \sum_{ v \in \mathcal{V}} \sum_{j \in \mathcal{L}, j \neq i} y_{i,j,v} = 1 , \forall i \in \mathcal{U}, \label{ilf-cons5} \\
	&& D^{\mathrm{dem}}_{i} \leq q_{i} \leq C^{\mathrm{cap}}_{v}, \forall i \in \mathcal{U}, \forall v \in \mathcal{V}, \label{ilf-cons6}  \\
	&&	 q_{i} \leq C^{\mathrm{cap}}_{v} + y_{0,i,v}( D^{\mathrm{dem}}_{i} - C^{\mathrm{cap}}_{v}), \forall v \in \mathcal{V}, \forall i \in \mathcal{U}, \label{ilf-cons7}\\
	&& q_{j} \geq q_{i} + D^{\mathrm{dem}}_{j} - C^{\mathrm{cap}}_{v} + (y_{i,j,v} C^{\mathrm{cap}}_{v}) \label{ilf-cons8} \\ \nonumber 
	&& + y_{j,i,v} (C^{\mathrm{cap}}_{v} - D^{\mathrm{dem}}_{j} - D^{\mathrm{dem}}_{i} ), \forall v \in \mathcal{V}, \forall i, j \in \mathcal{U}, i \neq j,  \\
	&& x_{v} = z_{0,v}, \forall v \in \mathcal{V},  \label{ilf-cons9} \\ 
	&& q_{i} \geq 0, \forall i \in \mathcal{U}, \label{ilf-cons10} \\ 
	&& z_{i,v} \in \{0, 1\}, \forall i \in \mathcal{U}, \forall v \in \mathcal{V}, \label{ilf-cons11} \\
	&&  x_{v}, y_{i,j,v} \in \{0, 1\}, \forall v \in \mathcal{V}, \forall i, j \in \mathcal{L}, i \neq j. \label{ilf-cons12}	 
\eeqn
\end{figure}

The objective function in (\ref{ilf-obj}) consists of three terms. The first term $\sum_{v \in \mathcal{V}} x_{v} C^{\mathrm{inc}}_{v}$ is the total initial cost. $C^{\mathrm{inc}}_{v}$ is the vehicle- and driver-based cost~\cite{aoct2015-analysis}. $x_{v}$ is the binary decision variable in which $x_{v} = 1$ if vehicle $v \in {\mathcal{V}}$ of the carrier is used to deliver packages, and $x_{v} = 0$ otherwise. The second term $\sum_{v \in \mathcal{V}} \sum_{i \in \mathcal{L}} \sum_{j \in \mathcal{L}, i \neq j} y_{i,j,v} C^{\mathrm{tvc}}_{i,j,v}$ is the total traveling cost. $C^{\mathrm{tvc}}_{i,j,v}$ is the traveling cost of vehicle $v$ moving from terminal $i$ to terminal $j$. This traveling cost is calculated from $C^{\mathrm{tvc}}_{i,j,v} = P^{\mathrm{ppm}} D^{\mathrm{dst}}_{i,j}$, where $P^{\mathrm{ppm}}$ is price per mile~\cite{aoct2015-analysis} and $D^{\mathrm{dst}}_{i,j}$ is a distance from terminal $i$ to terminal $j$. $y_{i,j,v}$ is the binary decision variable in which $y_{i,j,v} = 1$ if vehicle $v$ delivers packages to terminal $j$ after terminal $i$, and $y_{i,j,v} = 0$ otherwise. The third term $\sum_{v \in \mathcal{V}} \sum_{i \in \mathcal{U}} z_{i,v} C^{\mathrm{mcn}}$ is the total fixed cost. $C^{\mathrm{mcn}}$ is the miscellaneous cost of visiting a terminal by a vehicle, e.g., package handling cost. $z_{i,v}$ is the binary decision variable in which $z_{i,v}=1$ if vehicle $v$ delivers packages to terminal $i$, and $z_{i,v} = 0$ otherwise. 

The constraints are described as follows: (\ref{ilf-cons1}) and (\ref{ilf-cons2}) ensure that terminal $i$ is visited only once by vehicle $v$. For these constraints, the vehicle $v$ has to deliver all packages required by customers to terminal $i$~\cite{r-kaewpuang-cooperative-management2017},\cite{x-li-cvrpwtw-case-study2015} . (\ref{ilf-cons3}) guarantees that the number of vehicles used for package delivery does not exceed the number of available vehicles of a carrier. (\ref{ilf-cons4}) and (\ref{ilf-cons5}) guarantee that only one vehicle travels from terminal $i$ to terminal $j$. (\ref{ilf-cons6}) indicates that the delivery to terminal $i$ (i.e., $q_{i}$) has to meet the demand~\cite{r-kaewpuang-cooperative-management2017},\cite{x-li-cvrpwtw-case-study2015}. Additionally, the delivery must not exceed the capacity of vehicle $v$. The demand of terminal $i$ and the capacity of vehicle $v$ are denoted by $D^{\mathrm{dem}}_{i}$ and $C^{\mathrm{cap}}_{v}$, respectively. $q_{i}$ is the total number of packages to be delivered on the route that terminal $i$ is in. Note that, in practice, the customers' demands are packed into packages before delivery. In addition, the vehicle's capacity can be examined by the driver of the vehicle. If the vehicle's capacity is not enough for packages, an additional vehicle will handle packages. (\ref{ilf-cons7}) is to indicate whether or not terminal $i$ is the first terminal, the package of which is to be delivered on the route. (\ref{ilf-cons8}) applies, if terminal $i$ is not the first terminal of the route, to ensure that the number of delivered packages at terminal $j$ (i.e., $q_{j}$) must be more than or equal to the sum of the number of delivered packages on the route from the depot to terminal $i$ and the number of packages from terminal $j$ (i.e., $D^{\mathrm{dem}}_{j}$). (\ref{ilf-cons9}), (\ref{ilf-cons11}), and (\ref{ilf-cons12}) represent the binary integer constraints of the corresponding variables. (\ref{ilf-cons10}) is the non-negative integer constraint of the corresponding variables. 

The complexity of the IP model presented in (\ref{ilf-obj})-(\ref{ilf-cons12}) is $\mathcal{O}(|\mathcal{L}|^{2})$~\cite{j-k-lenstra-complexity-of-vr1981} where $\mathcal{O}(\cdot)$ and $|\mathcal{L}|$ are big $\mathcal{O}$ notation and the maximum number of terminals to be visited by the vehicles, respectively. This complexity means that the computational time of solving the IP model to achieve the solution increases exponentially when the number of terminals increases. 

The optimal solutions of (\ref{ilf-obj})-(\ref{ilf-cons12}) are $x_{v}^{\ast}, y_{i,j,v}^{\ast}$, and $z_{i,v}^{\ast}$ which yield the lowest delivery cost of the carrier. The traveling delay is then calculated as follows:
\begin{equation}
	\Phi ( \mathcal{L} ) = \sum_{ i,j \in \mathcal{L} } \frac{ D^{\mathrm{dst}}_{i,j} }{ V_{i,j,v}^{\mathrm{spd}}  } y_{i,j,v}^{\ast}	,
\label{eq:delay}
\end{equation}
where $V_{i,j,v}^{\mathrm{spd}}$ is the average speed of vehicle $v$ traveling from terminal $i$ to terminal $j$, and $D^{\mathrm{dst}}_{i,j}$ is the distance between terminals $i$ and $j$. This traveling delay of the carrier's vehicle is defined as the function of the set of terminal $\mathcal{L}$. The traveling delay is one of the factors that the customers are concerned about in selecting the carrier with which the stochastic evolutionary game to analyze the decision will be presented.

\section{Stochastic Evolutionary Game Formulation}
\label{sec:gameformulation}

In this section, we formulate the stochastic evolutionary game to model the carrier selection of customers. The game has four components as follows: {\bf Players} are the customers. {\bf Population} is the set of customers. {\bf Strategy} of customers is to select one of the LTL carriers. The set of carriers, i.e., strategy space, is denoted by ${\mathcal{C}}$. {\bf Negative payoff} is defined by its disutility which is a function of the service fee and average package delivery delay. The disutility of strategy $c \in {\mathcal{C}}$, i.e., a customer selects carrier $c$, is defined as follows:
\begin{equation}
	U_c ( n_c ) = F_c + \pi( n_c ),
\end{equation}
where $n_c$ is the number of customers selecting carrier $c$, $F_c$ is the service fee, and $\pi( n_c )$ is the average delivery delay of carrier $c$. By economic implication, this disutility implies that customers will obtain benefits (e.g., high reliability due to trustfulness and preference from terminals) by selecting appropriate carriers. In addition, carriers will achieve the minimum traveling cost due to balanced workloads.  

The average delivery delay given $n$ customers choosing the carrier is expressed as follows:
\begin{equation}
	 \pi( n ) =   n U +  \overbrace{\sum_{i_{1} = 1}^T \sum_{i_{2} = 1}^T \cdots \sum_{i_n = 1}^T }^n \left(  		\prod_{j=1}^T	\left(	\mathbb{P}_j \right)^{M_{j} } 	\Phi(\mathcal{L}_c	(\{i_1,\ldots,i_n\}	)		\right)	\label{eq:util-obj}	
\end{equation}
where
\begin{eqnarray}
	M_{j} 					&	=	&	\sum_{s = 1, i_s = j}^n 1, 		\quad	\quad	j \in \mathcal{T},			\\
	\mathcal{L}_c	(\{i_1,\ldots,i_n \}	)			&	=	&	\Big\{ \{j\} \Big| j=1,\ldots,T ; M_j > 0	\Big\}.
\end{eqnarray}

$U$ is the package unloading time per customer, and hence $n U$ is the total unloading time for $n$ customers. In this paper, we assume that the package unloading time is known in advance from evaluating statistical information~\cite{ilogistics2016}. $i_s$ is the terminal to which the package of customer $s$ is to be delivered. $M_{j}$ is the number of customers, the packages of which are to be delivered to terminal $j$. $\Phi(\mathcal{L}_c)$ is the delivery delay of carrier $c$ given the set of terminals to visit $\mathcal{L}_c$ as it is obtained from (\ref{eq:delay}) by solving vehicle routing optimization defined in (\ref{ilf-obj})-(\ref{ilf-cons12}). $\mathbb{P}_j$ is the probability that the customer wants to deliver a package to terminal $j$.

In the game, each customer compares its disutility with the average disutility of all carriers. The customer switches to the new carrier if its current disutility is higher than that of the new carrier. With this strategic action of the customer population, we can model the strategic evolution of the customers using a Markov chain. The state space of the Markov chain is defined as follows:
\begin{equation}
	\Omega	=	\Big\{ (n_1,\ldots,n_N)	\Big| \sum_{c=1}^N n_c = S 	\Big\},
\end{equation}
where again $n_c$ is the number of customers selecting carrier $c$. To deal with the large state space of the Markov chain, the most probable state generation algorithm and techniques of decomposition, lumping, and truncation can be applied~\cite{e-de-souza-state-space1992}. The transition rate from state $(n_1,\ldots,n_N)$ to state $(n'_1,\ldots,n'_N)$ is expressed as follows:
\begin{equation}
	Q_{(n_1,\ldots,n_N),(n'_1,\ldots,n'_N)} = \left\{ \begin{array}{ll}
			K n_c ( U_c - U_{c'} )	,	&	\mbox{if } U_c > U_{c'}	,	\\
			\epsilon			,	&	\mbox{otherwise}	,
		\end{array}	\right.
\label{eq:transrate}
\end{equation}
where $K$ is a certain constant related to the decision-making rate of customers, and $\epsilon$ is a small perturbation rate related to the irrational decision-making of the players. In particular, the customer can make a small mistake to choose the carrier which yields higher disutility. From (\ref{eq:transrate}), the transition rate depends on the difference between the disutilities of two carriers, and the number of customers currently choosing an inferior carrier, i.e., with larger disutility. Note that the self-transition rate of each state~\cite{r-kaewpuang-cooperative-management2017} is expressed as follows: 
\begin{eqnarray}
	& & Q_{(n_1,\ldots,n_N),(n_1,\ldots,n_N)} = 	\nonumber	\\
	& & \quad \;\; - \sum_{	(n'_1,\ldots,n'_N) \neq (n_1,\ldots,n_N) } Q_{(n_1,\ldots,n_N),(n'_1,\ldots,n'_N)}	.	
\end{eqnarray}

Let ${\mathbf{Q}}$ denote the transition matrix of the Markov chain, the element of which is $Q_{(n_1,\ldots,n_N),(n'_1,\ldots,n'_N)}$, the stationary probability of state $(n_1,\ldots,n_N)$ of the Markov chain can be obtained by solving $\vec{\boldsymbol{\beta}}^\top {\mathbf{Q}} = \vec{\mathbf{0}}^\top$ and $\vec{\boldsymbol{\beta}}^\top \vec{\mathbf{1}} = 1$, where $\vec{\boldsymbol{\beta}}$ is the vector of stationary probability, $\vec{\mathbf{0}}$ is the vector of zeros, and $\vec{\mathbf{1}}$ is the vector of ones. For $\epsilon \rightarrow 0$, if the stationary probability of state $(n^\dag_1,\ldots,n^\dag_N)$ does not approach zero, then the state is stochastically stable. In other words, we can consider the state $(n^\dag_1,\ldots,n^\dag_N)$ to be a stable point of the game, i.e., an equilibrium.


\section{Performance Evaluation}
\label{sec:performanceevaluation}

\subsection{Parameter Setting}

To simplify the performance of our proposed approach, we consider the carrier selection of 16 customers. There are two LTL carriers, and three terminals (e.g., an airport, train station, and bus terminal). Note that our proposed approach can handle many customers (e.g., 100 customers) and terminals (e.g., 50 terminals). The locations of the two carriers and three terminals are shown in Fig.~\ref{fig:systemmodel-vp-c2-w-dist}(b) in which the unit is in kilometer. We adopt the data provided by I.logistics Pte Ltd company in Singapore~\cite{ilogistics2016}. Carriers 1 and 2 have one LTL vehicle. The vehicle capacities of carriers 1 and 2 are 20 and 30 packages, respectively. The package unloading time is 5 minutes per customer. The average speed of the vehicles is 40 km/h~\cite{ilogistics2016}. We implement and solve the vehicle routing optimization model of each LTL carrier using the GAMS/CPLEX solver~\cite{Gams}. The traveling cost of the LTL vehicle is 0.982\$ per kilometer~\cite{aoct2015-analysis}. We assume that all the carriers charge the same fee due to market competition. Thus, the customers focus only on the average delivery delay. For the Markov chain, we set $K=1$ and $\epsilon = 10^{-3}$.

\begin{table}[ht] 
\captionsetup{justification=centering}
\caption{Traveling delay of carriers 1 and 2 to different sets of terminals. }
\label{table:delay-ltl-1-2}
\centering
\captionsetup{justification=centering}
\scalebox{0.9}{\begin{tabular}{|c|c|c|c|}\hline
&		\multicolumn{2}{|c|}{\bf{Traveling delay (minutes)}}  \\\hline
Terminal & Carrier 1 & Carrier 2  			        \\ \hline
1 								&2.400 &18.228	    \\\hline 
2     							&8.625 &17.376	    \\\hline
3      							&9.450 &17.340		\\\hline
$1 \rightarrow 2 $  			&13.425 &29.253	    \\\hline
$1\rightarrow 3 $     			&14.250 &30.078		\\\hline	
$2\rightarrow 3 $   			&26.700 &35.415		\\\hline	
$1 \rightarrow 2 \rightarrow3$  &31.500 &47.328		\\\hline	
\end{tabular}}
\end{table}

Table~\ref{table:delay-ltl-1-2} shows the traveling delay of each carrier visiting different sets of terminals. Clearly, the traveling delay of carrier 1 is smaller than those of carrier 2 due to carrier 1's location which is closer to terminals. 

\begin{figure*}[t]
   \centering
   \captionsetup{justification=centering}
   \subfloat[Average delay without rational decision.]{\label{fig:irrat-selection}\includegraphics[width=0.2\textwidth]{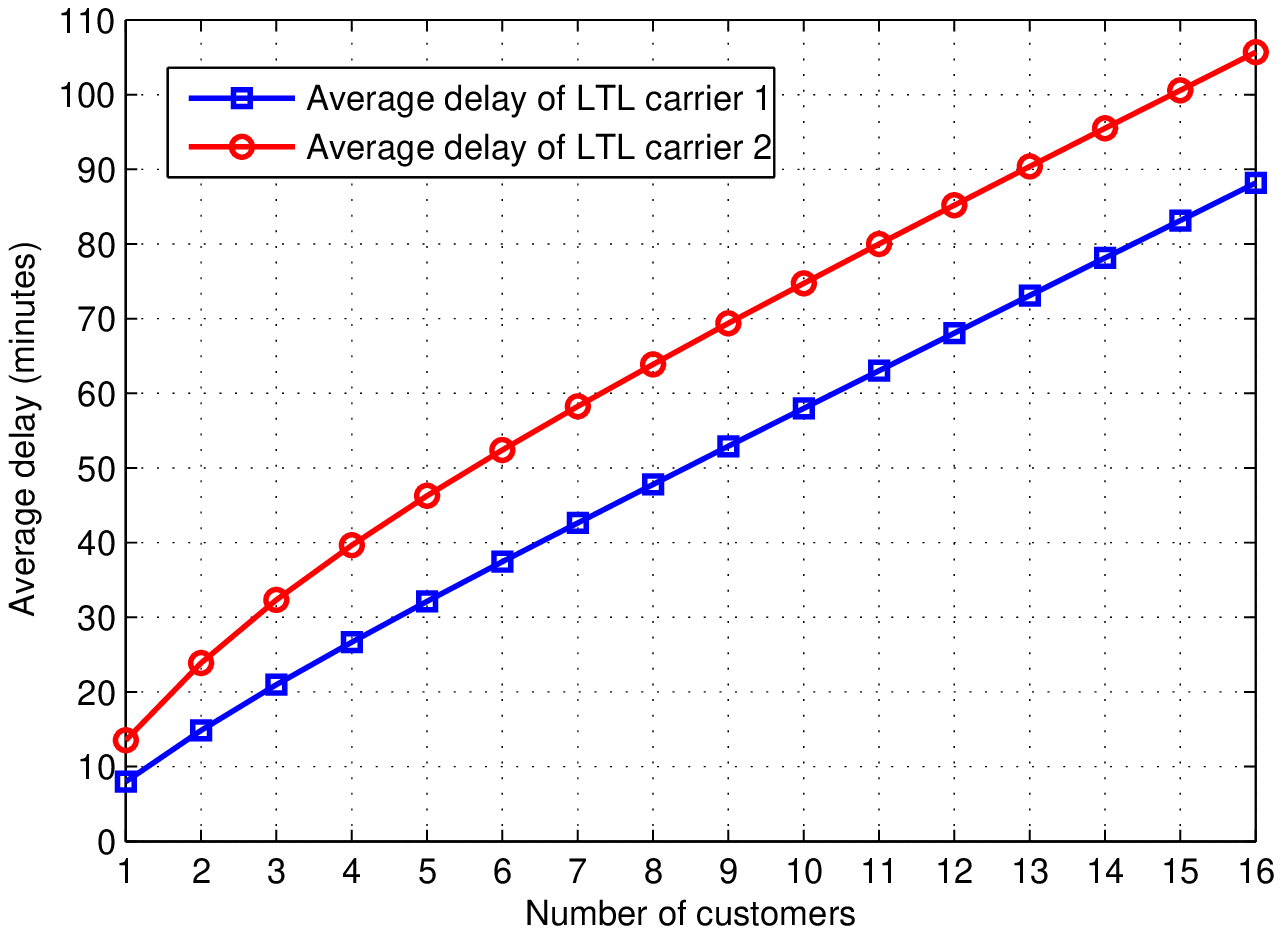}}
   \subfloat[Average delay under different  customers.]{\label{fig:avg-delay}\includegraphics[width=0.2\textwidth]{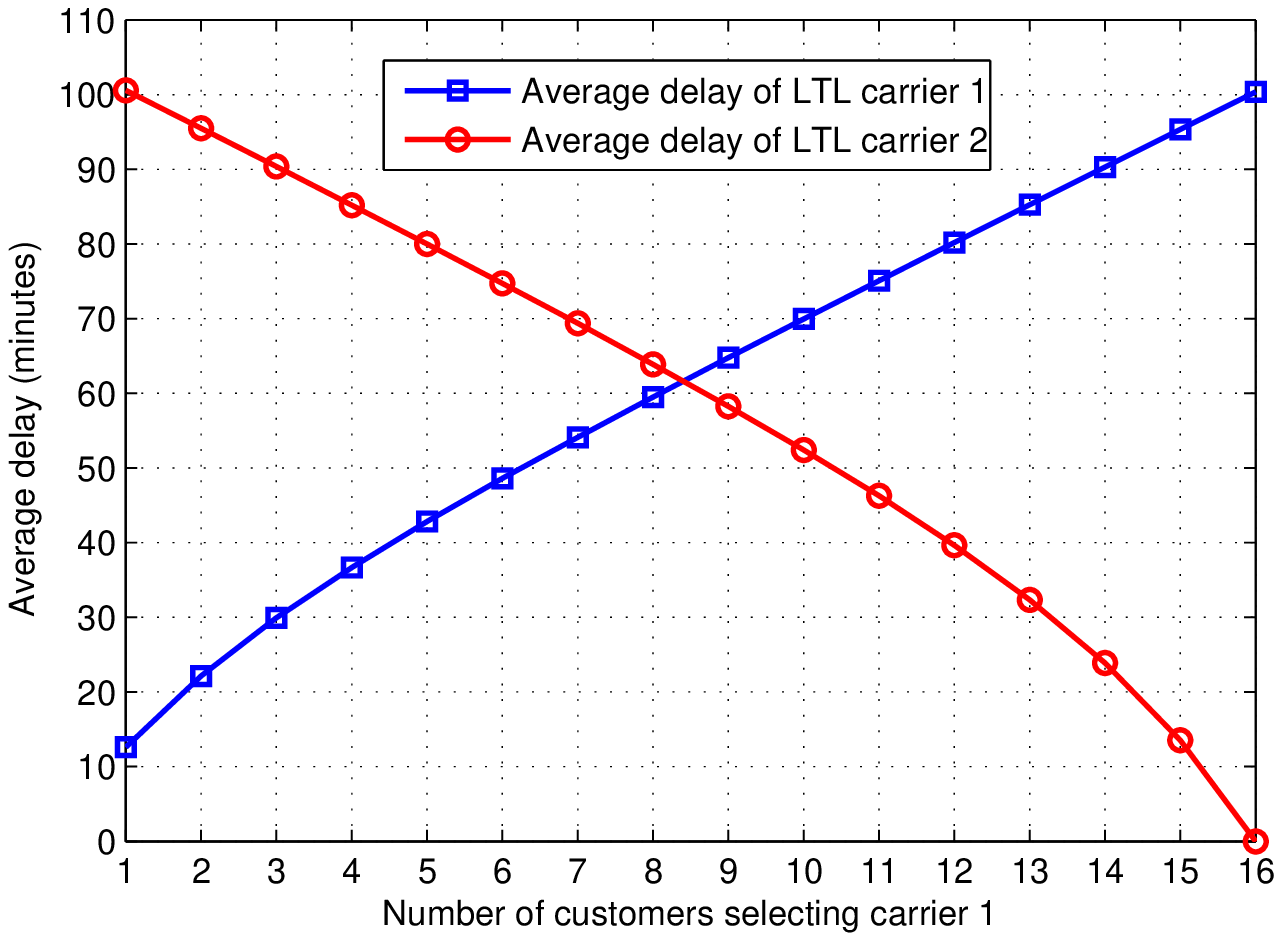}}
   \subfloat[Stochastically stable states of the evolutionary game.]{\label{fig:compareCost}\includegraphics[width=0.2\textwidth]{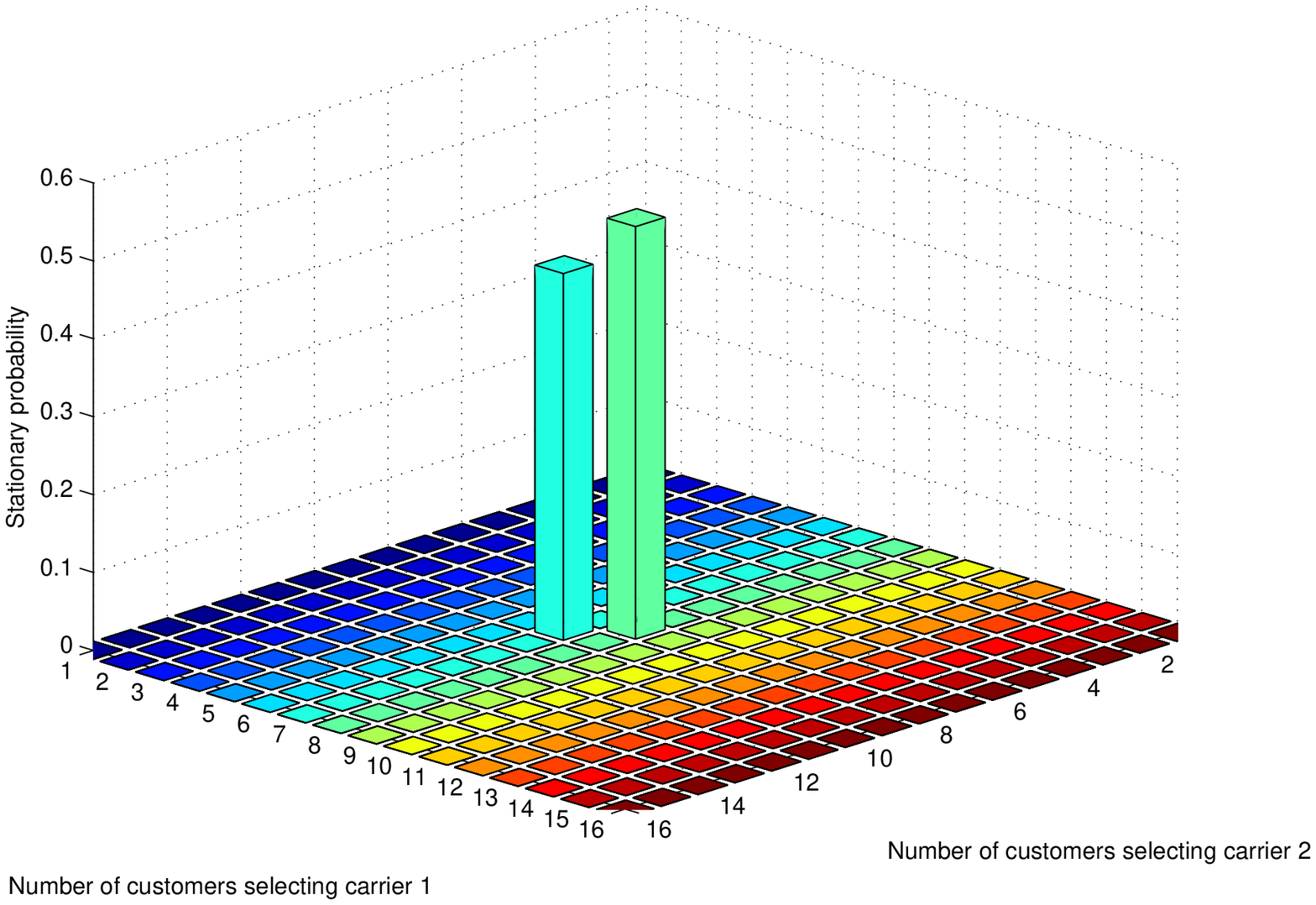}}
   \subfloat[Average number of customers choosing different carriers.]{\label{fig:compareCost}\includegraphics[width=0.2\textwidth]{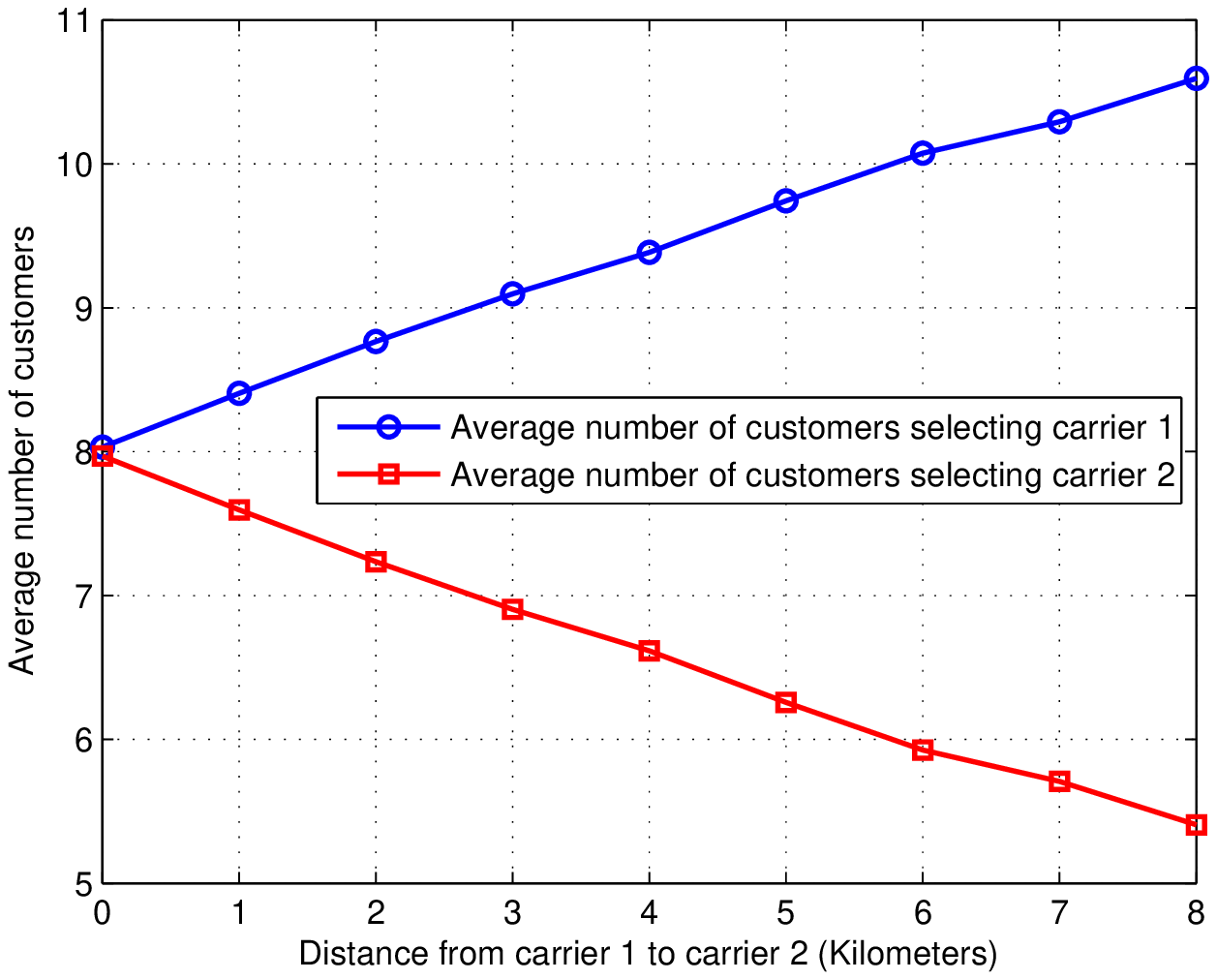}}
   \subfloat[Total traveling cost comparison of carrier 1]{\label{fig:compareCost}\includegraphics[width=0.2\textwidth]{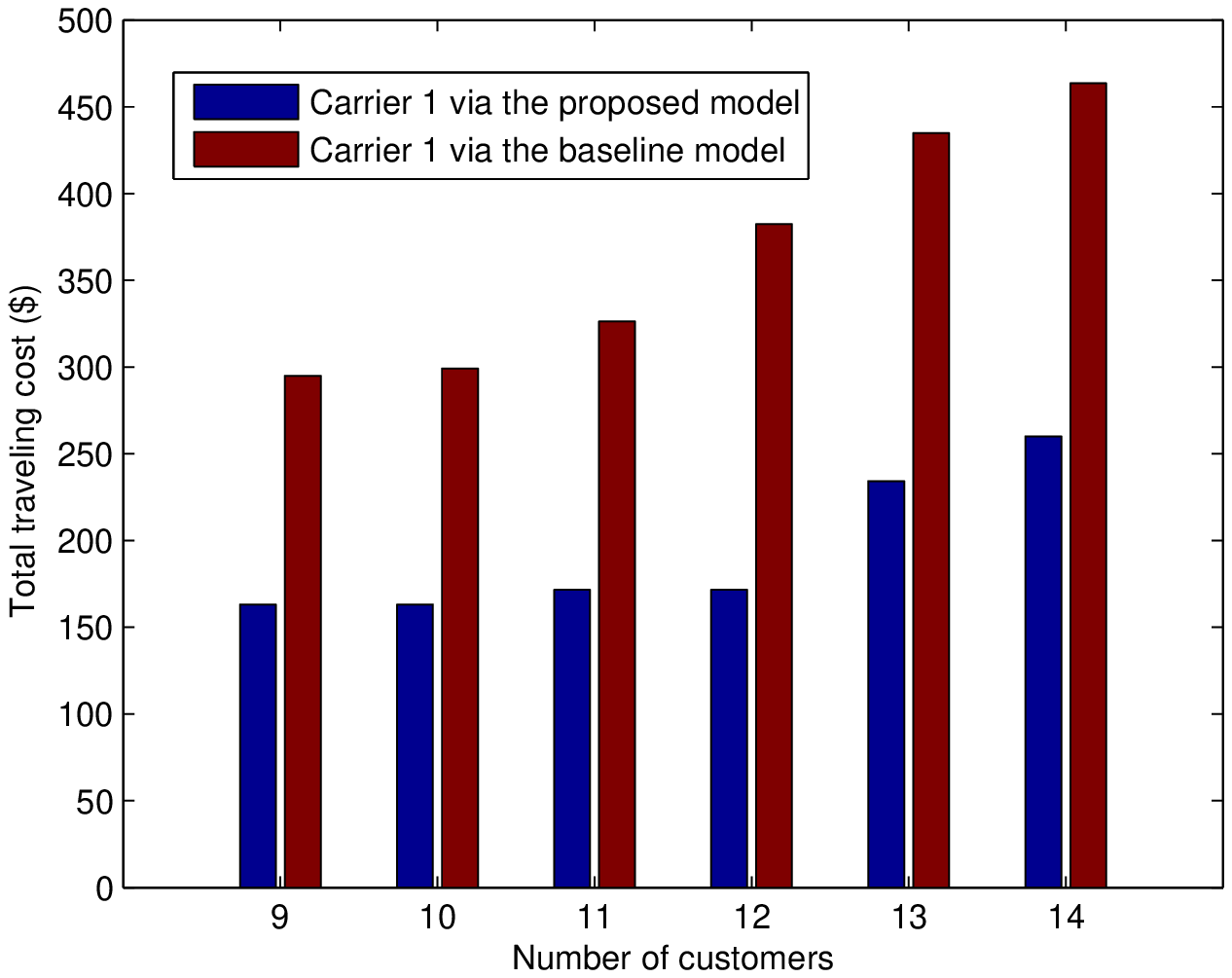}}
    \caption{(a) Average delay of carriers 1 and 2 without rational decision-making of customers, (b) Average delay of carriers 1 and 2 under the different number of customers choosing carrier 1, (c) Stochastically stable states of the evolutionary game, (d) The average number of customers choosing different carriers when the distance from carrier 2 to carrier 1 is varied, and (e) The total traveling cost comparison between the proposed model and the baseline model.}
   \label{fig:results}
\end{figure*}


Fig.~\ref{fig:results}(a) shows the average delay of carriers under different carrier locations when all the customers irrationally select either carrier 1 or carrier 2 to deliver packages to terminals. In other words, all the customers only select carrier 1 or carrier 2 since they do not consider the performance of carriers. In Fig. \ref{fig:results}(a), the average delay of carriers for package delivery dramatically increases when the number of customers increases. This is due to the fact that all the customers make the irrational decision to only select carrier 1 or carrier 2, and consequently the average delay of the carrier to deliver customers' packages to the terminals increases. Additionally, the total package unloading time of the carrier increases. 

From the result shown in Fig.~\ref{fig:results}(a), this is a reason why our proposed approach based on the evolutionary game is beneficial to analyze the carrier selection problem to minimize the average delay of carriers for package delivery.
 

We next consider the population of customers. Fig.~\ref{fig:results}(b) shows the average delivery delay from both carriers when the number of customers choosing the carriers is varied. For example, if the number of customers choosing carrier 1 is 2, then the number of customers choosing carrier 2 will be 14. From Fig.~\ref{fig:results}(b), when the number of customers choosing each carrier increases, the average delivery delay increases. This is due to the fact that, with more customers, the chance that the carrier has to visit more terminals increases. Additionally, the total package unloading time increases. Therefore, the decision of the customer to choose one carrier to minimize its disutility also depends on the decisions of other customers. It is important to analyze the decision-making of the customers and obtain stable states of the carrier selection.

From the evolutionary game for the carrier selected by customers, the stochastically stable states, i.e., the number of customers choosing carriers, are shown in Fig.~\ref{fig:results}(c). We make the following observations. Firstly, stochastically stable states have the non-zero stationary probability as $\epsilon \rightarrow 0$. Secondly, the stochastically stable states associate with the number of customers choosing the carriers that result in a similar average delivery delay. Thirdly, there are multiple stochastically stable states. This is due to the fact that these states result in slightly higher or lower disutility for the customers. Therefore, the customers can switch among the states to achieve identical average disutility. In stochastically stable states shown in Fig.~\ref{fig:results}(c), the first stable state means that eight customers select carrier 1 and eight customers select carrier 2. The second stable state means that seven customers select carrier 1 and nine customers select carrier 2.  

Next, we evaluate the impact of the location of a carrier on the decisions of customers. We assume that initially carrier 2 is at the same location as carrier 1 (see Fig.~\ref{fig:systemmodel-vp-c2-w-dist}(b)). Then, the location of carrier 2 is shifted to the east (the right side) of carrier 1 horizontally, which implies that carrier 2 becomes farther away from the terminals. Fig.~\ref{fig:results}(d) shows the average number of customers choosing the carriers when the distance from carrier 1 to carrier 2 is varied. Clearly, as the distance becomes larger, carrier 2 takes more time to reach all three terminals, and hence the delivery delay increases. Consequently, due to poorer performance, customers deviate to choose carrier 1. This result shows evidently the usefulness of the proposed stochastic evolutionary game model that can capture the physical parameters of the vehicle routing problems. 

Finally, we compare the traveling costs of carrier 1 from the proposed model and the baseline model. Figure~\ref{fig:results}(e) shows the performance comparison of the proposed model and the baseline model by varying the number of customers. For the baseline model, we consider that all customers make the irrational decision to select carrier 1 only. Clearly, the proposed model can achieve a significantly lower traveling cost of carrier 1 than the baseline model.

\section{Conclusion}
\label{sec:conclusion}

Due to limited resources, i.e., vehicles of carriers, package delivery performance can depend on the number of customers served by the carrier for package delivery. In this paper, we have considered the less-than-truckload services so that the customers can select the carrier with the best performance. As the customers can adapt their decision dynamically, we have presented the stochastic evolutionary game model to analyze the decision-making of the population of customers. We have first formulated vehicle routing optimization for obtaining optimal delivery routes to minimize the cost of a carrier. We have then formulated the evolutionary game model to obtain the stochastically stable states in terms of the number of customers choosing different carriers. The disutility of the customers in the evolutionary game is a function of delivery delay obtained from the vehicle routing optimization problem. The numerical results have verified the existence of the stochastically stable states for which all customers receive identical average disutility. Interestingly, in one of the stochastically stable states, seven customers select carrier 1 while nine customers select carrier 2 due to the identical average disutility. 

For future work, we will consider multiple populations of customers, the (dis)utility of which are defined differently. The service fee can be optimized to maximize the profits of the carriers. In addition, we will consider the effect of the average speed of the vehicle on the equilibrium solution. 
   
\section*{Acknowledgment}
This research is supported by the National Research Foundation Singapore and DSO National Laboratories under the AI Singapore Programme (AISG Award No: AISG2-RP-2020-019); the National Research Foundation (NRF), Singapore and Infocomm Media Development Authority under the Future Communications Research Development Programme (FCP); Energy Research Test-Bed and Industry Partnership Funding Initiative, part of the Energy Grid (EG) 2.0 programme; DesCartes and the Campus for Research Excellence and Technological Enterprise (CREATE) programme; Alibaba Group through Alibaba Innovative Research (AIR) Program and Alibaba-NTU Singapore Joint Research Institute (JRI); and Nanyang Technological University, Nanyang Assistant Professorship.

\end{document}